 \documentclass[aps,pra, onecolumn,14 pts,a4paper,showpacs]{revtex4}

\usepackage{amssymb,amsmath,amsfonts,graphicx}
\usepackage{bm}

\begin{document}

\title{Fluid motion induced by surface waves at low Reynolds number}

\author{Yves Pomeau}
\affiliation{Department of Mathematics, University of Arizona, Tucson, AZ USA}

\date{\today }
\begin{abstract}
\textbf{Abstract}
We discuss the scaling laws for the flow generated in a viscous fluid by a wave propagating along a solid boundary. This has applications to the displacement of tiny objects on solids, under the effect of progressive surface waves and for the swimming of microanimals by undulation of ciliae along their body surface

\end{abstract}

\maketitle
%\section{Introduction}
 
 Suppose a channel with an undulation on one side of amplitude $h$, representing the propagation of a surface wave of phase speed $c$. Therefore the fluid velocity induced in the channel must be quadratic with respect to $h$. This fluid speed is induced near the undulating surface, and so it should be first along the surface itself and then depending on the local properties of the surface. The only combination giving the velocity induced by this wave motion is $$u \sim (\nabla h)^2 c \mathrm{,}$$ where $\nabla h$ is the 2D gradient of the wave amplitude as a function of the coordinate in the plane of the plate. In an eventual realisation with Leidenfrost  droplets hovering at an elevation $H$, as derived in \cite{CRAS}, $H$ being the height iof the channel underneath the vapor bubble, the tangential force due to this flow is $F = \eta  \frac{(\nabla h)^2}{H}  c\  r_c^2$, where $r_c$ is the radius of the area flattened underneath the levitating droplet. This formula is interesting because, although the slope of the surface induced by the wave, $\nabla h$, is likely quite small, the speed of surface waves \cite{LL} in usual solids is quite large, in the range of thousand meters per second. With a somewhat arbitrary $\nabla h$ of order $10^{-3}$ this yields $u$ about a millimeter per second, not so small. 
 
 The formula given above for the velocity induced by the progressing wave is simple to derive. Let $u$ be the horizontal component of the velocity and $w$ be its vertical component, normal to the average position of the plane limiting te solid where the surface wave propagates. Let ${\bf{u}} = (u, w)$ be the velocity vector. The shape of the surface is given by the Cartesian equation 
 $$ z = h(x -ct)$$ 
 where $c$ is the speed of the wave. Because we consider small scale phenomena, the fluid must be seen as viscous and the boundary conditions on the surface of the solid (namely $ z = h(x -ct)$) are:
 $$ {\bf{u}}\cdot {\bf{t}} = 0 \mathrm{,}$$ and 
 $${ \bf{u}} \cdot {\bf{n}} ={ \bf{U}} \cdot {\bf{n}} \mathrm{.}$$
 In the boundary conditions, $\bf{n}$ and $\bf{t}$ are vectors tangent and normal to the surface, ${\bf{u}} =  (u, w)$ is the fluid velocity and $\bf{U}$ is the velocity of the solid surface.  The point on the solid surface has coordinates $(x, z = h(x -ct)$. Therefore ${\bf{U}} = (0, - c h_{,x})$ where $h_{,x} = \frac{\mathrm{d} h}{\mathrm{d} x}$. Moreover one can take (no normalisation is needed) ${\bf{t}} = (1, h_{,x})$ and ${\bf{n}} =  (- h_{,x}, 1)$. Therefore the two boundary conditions become 
 $$ u + w h_{,x} = 0 $$ and $$ w = ( u - c) h_{,x}$$ 
 This yields after simple algebra $$ u = c \frac{h_{,x}^2}{! + h_{,x}^2}$$ 
 Therefore, for a weak amplitude wave, $$ u \approx c h_{,x}^2$$ 
 This makes a boundary condition on the plane surface where the surface wave propagate. Therefore on this surface one has to introduce in the boundary conditions for the fluid equations a constant tangential velocity given by the time average of $h_{,x}^2$ defined as $<h_{,x}^2>$. Supposing a geometry of a simple parallel channel, one obtains a standard Couette flow with a velocity depending linearly on $z$. If the other side of the channel is at rest, one finds the estimate:
 $$ u \approx c <h_{,x}^2> (1 - \frac{z}{H}) $$ equivalent to the formula given before whenever the wavelength of the wave is of the same order of magnitude as the width of the channel. It shows also that, for given $H$ and $c$, the phenomenon is the more important if the average slope of the surface waves is the bigger.  The general formula valid for the horizontal force on a Leidenfrost droplet is  
 $$ F \sim \eta r_c^2 \frac{<h_{,x}^2>}{H} $$

Let us notice also that this creation of a mean flow by undulations on a boundary is rather common in biology \cite{RevBio}: many tiny animals swim (at low Reynolds number) by undulating layers of ciliae on their body. Compared to surface waves in solids there, presumably, the equivalent of the slope $\nabla h$ may be quite large of order one but, the celerity $c$ is much smaller! 
      {}

      \ifx\mainismaster\UnDef%
      \end{document}
      \fi